\documentclass[draft]{agujournal2019}
\usepackage{url} 
\usepackage[inline]{trackchanges} 
\usepackage{soul}

\draftfalse

\journalname{JGR: Atmospheres}

\begin{document}

\title{Medium-scale thermospheric gravity waves in the high-resolution Whole Atmosphere Model: Seasonal, local time, and longitudinal variations}

\authors{Garima Malhotra\affil{1}, Timothy Fuller-Rowell\affil{1,2}, Tzu-Wei Fang \affil{2}, Valery Yudin\affil{3}, Svetlana Karol\affil{1,2}, Erich Becker\affil{4}, Adam Marshall Kubaryk\affil{1,2}}

\affiliation{1}{CIRES, University of Colorado Boulder, CO}
\affiliation{2}{Space Weather Prediction Center, National Oceanic and Atmospheric Administration, CO}
\affiliation{3}{Catholic University of America, DC}
\affiliation{4}{Northwest Research Associates, CO}

\correspondingauthor{Garima Malhotra}{garima.malhotra@colorado.edu}

\begin{keypoints}
\item WAMT254 GWs (in zonal winds; scales$<$620 km) can reasonably reproduce observational GW climatology in the lower atmosphere and thermosphere.

\item WAM GWs propagate up from MLT into the IT in the westward phase of migrating tidal winds and thus show diurnal and semidiurnal variations.

\item WAM predicts a wave-4 signature in daytime tropical thermospheric GW activity due to modulation by the non-migrating tides in the MLT.

\end{keypoints}

\begin{abstract}
This paper presents a study of the global medium-scale (scales$<$620 km) gravity wave (GW) activity (in terms of zonal wind variance) and its seasonal, local time and longitudinal variations by employing the enhanced-resolution ($\sim$50 km) Whole Atmosphere Model (WAMT254) and space-based observations for geomagnetically quiet conditions. It is found that the GW hotspots produced by WAMT254 in the troposphere and stratosphere agree well with previously well-studied orographic and non-orographic sources. In the ionosphere-thermosphere (IT) region, GWs spread out forming latitudinal band-like hotspots. During solstices, a primary maximum in GW activity is observed in WAMT254 and GOCE over winter mid-high latitudes, likely associated with higher-order waves with primary sources in polar night jet, fronts and polar vortex. During all the seasons, the enhancement of GWs around the geomagnetic poles as observed by GOCE (at $\sim$250 km) is well captured by simulations. WAMT254 GWs in the IT region also show dependence on local time due to their interaction with migrating tides leading to diurnal and semidiurnal variations. The GWs are more likely to propagate up from the MLT region during westward/weakly-eastward phase of thermospheric tides, signifying the dominance of eastward GW momentum flux in the MLT. Additionally, as a novel finding, a wavenumber-4 signature in GW activity is predicted by WAMT254 between 6-12 LT in the tropics at $\sim$250 km, which propagates eastward with local time. This behavior is likely associated with the modulation of GWs by wave-4 signal of non-migrating tides in the lower thermospheric zonal winds. 

\end{abstract}

\section*{Plain Language Summary}
Gravity waves (GWs) are the buoyancy-driven waves in the Earth's atmosphere with periods ranging from few minutes to a few hours. These waves are generated by various processes and play an important role in coupling different regions and layers of the atmosphere. In this study, we use a global high-resolution whole atmosphere model that spans the Earth's atmosphere from the surface to the thermospheric and ionospheric heights of $\sim$600 km. We investigate the seasonal climatology of GWs in the troposphere, stratosphere and thermosphere generated in the model and find that it agrees with previous knowledge from observations and models. We also find that the thermospheric GW activity in the model depends on local time and longitude signifying its interaction with atmospheric tides. Specifically, we observe a wavenumber-4 in the crests and troughs of GW activity around the equator during daytime, which is a signature of non-migrating tides modulating the propagation of GWs. This hasn't been observed before and elucidates the importance of high-resolution whole atmosphere models in advancing the knowledge of upper atmospheric physics.

\section{Introduction}
The variability in the Ionosphere and Thermosphere (IT) is largely driven by the solar and geomagnetic activity, but the dynamic coupling with the lower and middle atmosphere through various waves (e.g., tides, traveling planetary waves, and gravity waves) also plays an important role. Lower atmospheric waves can affect the background state of the mesosphere and lower thermosphere (MLT) region which can then impact the dynamics and state of the IT region \cite{Qian09, Jones17, Malhotra20, Malhotra22}. Many of the upper atmospheric phenomena also demonstrate the importance of lower atmospheric gravity waves (GWs) for the MLT and IT region, e.g., mesospheric residual circulation and the resulting cold summer mesopause \cite{Houghton78, Lindzen81, Holton82}, lower thermospheric residual circulation \cite{Qian17a, Qian17b}, mesospheric quasi biennial oscillation \cite{Burrage96, Malhotra16}, and ionospheric scintillations and plasma bubbles \cite <e.g.,>{Hysell90, Kelley81, Fritts08, Wu21}. Moreover, some global circulation models suggest that the gravity wave drag is comparable to the ion drag in the momentum budget above the turbopause \cite{Yigit09b, Yigit10, Yigit12a, Miyoshi14}.
 
Atmospheric GWs are primarily generated in the lower atmosphere from various phenomena such as wind shears \cite{Buhler99, Scinocca00}, flow over topography \cite <e.g.,>{Nastrom87, Nastrom92}, convective activity \cite <e.g.,>{Clark86, Mclandress00}, geostrophic adjustment \cite{Fritts92, Vadas01}. They play an important role in coupling the energy and momentum between different regions of the Earth's atmosphere \cite{Houghton78}. During geomagnetic active times, gravity waves can also be generated in-situ in the IT region at auroral latitudes \cite{Bruinsma08}. Natural hazards such as the earthquakes, tsunamis and volcanoes can also launch GWs that can reach IT heights \cite<e.g.,>{Kelley85, Artru05, Coisson15, Wright22}. In addition, GWs can be generated in the winter stratosphere via spontaneous emission from the polar vortex \cite{Sato08, Becker22b, Vadas23}. Most importantly, the generation of secondary and higher-order GWs in the middle and upper atmosphere has been found to be a crucial source for of thermospheric GWs \cite{Becker18, Vadas18, Vadas19, Becker20, Becker22}.

GWs propagate both horizontally and vertically away from their source regions with horizontal wavelengths ranging from tens to thousands of kilometers (kms) \cite<e.g.,>{Vincent00, Ern04, Wang06, Preusse08}. Their vertical wavelengths change significantly with altitude from tens of kms in lower thermosphere to hundreds of kms at higher altitudes in the thermosphere \cite{Vadas07}. Their intrinsic periods range from a few minutes (buoyancy period) to hours (inertial period) with waves of smaller periods and larger phase velocities dominating at higher altitudes \cite{Vadas07, Fritts08b, Miyoshi08}. GWs dissipate depending on the background conditions, thereby depositing their momentum and energy and giving rise to changes of the background flow. The mixing and diffusion due to breaking of gravity waves can also change the composition of the IT region \cite{Hodges67, Qian09, Malhotra17, Malhotra21}. Moreover, the wide spectrum of secondary and higher-order waves generated by breaking of GWs can propagate further up into the IT region and contribute to its variability \cite{Satomura99, Vadas02, Vadas17, Becker22}. 

GW activity at different altitudes has been reported by numerous studies using ground, balloon and space-based instruments \cite<e.g.,>[and others]{Ern04, Preusse08, Preusse09, John12,  Park14, Walterscheid16a, Walterscheid16b, Liu17, Liu17a}. Below the MLT, different satellite instruments have been employed to capture different portions of the GW spectrum, often showing contrasting results. For example, \citeA{Tsuda00} using Global Positioning System (GPS) temperature profiles observed an equatorial maximum in the stratospheric GW activity during Nov-Feb. Conversely, \citeA{Wu96} using Microwave Limb Sounder (MLS) radiances reported a minimum in GW variances at the equator This difference was later studied to be due to the fact that the GPS analysis method is sensitive to GWs with short vertical wavelengths, while MLS analysis favors GWs with longer vertical wavelengths \cite{Alexander07}. Furthermore, there is still an observational gap for GWs with smaller vertical and horizontal wavelengths in the lower atmosphere \cite{Preusse08}. On the other hand, in the IT region, most of the GW observations come from electron density measurements and airglow emissions, which are in principle tracers of GW perturbations in the thermosphere \cite{Liu06, Hickey10, Fukushima12}. In the middle and upper thermosphere, the limited global measurements of GW perturbations mostly come from in-situ accelerometer measurements by GOCE at 250 km \cite{Forbes16, Liu17}, GRACE, GRACE-FO \cite{Park23}, and CHAMP \cite{Bruinsma07, Bruinsma08, Park14} between $\sim$400-500 km. Because of their wide spatial and temporal scales, a comprehensive understanding of the GW spectrum, propagation, dissipation in the IT region and their link to the aforementioned various sources is still lacking. A deeper understanding of direct link between GW signatures of different altitude regimes would require simultaneous routine observations of GWs in the lower and upper thermosphere which are quite hard to achieve and have only recently gained attention. 

To better understand the seamless vertical coupling due to GWs from the lower to the upper atmosphere without introducing artificial atmospheric boundaries, whole atmosphere models were developed \cite{Roble00, Akmaev11, Liu18}. The resolution required to fully resolve the wide spectrum of GWs in such models is orders of magnitude higher than what has been computationally affordable in the past. Therefore, using the GW observations, sub-grid parameterizations were developed to model the effects of upward propagating GWs on the large-scale momentum, composition and temperature of the MLT region \cite<e.g.,>{Lindzen81, Fritts93, Lott97, Hines97a, Hines97b, Alexander99, Becker09}. These parameterizations have been widely successful in reproducing GW effects in the middle atmosphere. One of the major limitations of parameterizations is the bias and uncertainty due to gaps in GW observations, many tunable parameters, and oversimplification of GW characteristics (e.g., GW sources, assumption of single column vertical propagation, steady-state approximation, no secondary wave generation)  \cite{Alexander10, Camara14, Plougonven20, Wei22}. Some IT models with boundaries in the MLT region, e.g., GITM, TIEGCM \cite{Dickinson81, Ridley06} use a simplistic eddy diffusion parameter to include the effects of gravity wave mixing in the lower thermosphere \cite{Qian09, Malhotra22}, which fails to include non-local effects from GWs on momentum and energy deposition. 

With recent advances in supercomputing, higher resolution whole atmosphere models are able to better resolve the mesoscale portion of the GW spectra. The middle atmosphere circulation is especially sensitive to the horizontal resolution of a GW-resolving model \cite{Hamilton99}, which can impact the propagation of GWs into the IT region. Several studies have demonstrated that with high model resolution, several effects of GWs on the upper atmosphere can be self-consistently reproduced without GW parameterizations  \cite<e.g.,> {Hayashi89, Watanabe08, Geller13, Liu14c, Miyoshi14, Becker20}. Enhancing the resolution of a whole atmosphere model not only provides a realistic background for GWs that propagate from the lower atmosphere into the IT region but allows in addition for the simulations of secondary and higher-order GWs which play a crucial role in the IT \cite<e.g.>{Vadas19, Becker20, Becker22}. 

In this study, we use the state-of-the-art high resolution (0.5$^{\circ}$ horizontal resolution) global spectral whole atmosphere model (GSMWAM or WAM) to investigate the latitudinal and seasonal climatology of medium-scale GW activity in the lower and upper thermosphere and its correlations with the variability of the lower atmosphere. Many studies have demonstrated that WAM can realistically simulate the midnight temperature and density maxima \cite{Akmaev09, Akmaev10}, the wave-4 longitude structure of ionospheric electrodynamics \cite{Fang13}, and the response of the upper atmosphere to sudden stratospheric warming \cite{Fuller10}. \citeA{Yudin17} and \citeA{Yudin22a, Yudin22b} reported that multi-year WAM simulations constrained by Goddard Earth Observing System (GEOS) meteorology \cite{Rienecker08} reproduce the day-to-day, seasonal, and year-to-year variability of tidal dynamics in the upper atmosphere. 

We define medium-scale GWs as having horizontal wavelengths ranging between 200-620 km. This definition is based on previous studies that observed the thermospheric GWs with scale sizes $<$600 km \cite{Forbes95, Bruinsma08, Forbes16, Garcia16, Liu17, Park14}. This definition excludes most of the larger-scale GWs generated by the geomagnetic activity or signatures of equatorial thermosphere mass density anomaly, while capturing a large proportion of GWs generated by jets, fronts and flow over orography \cite{Preusse08, Plougonven16}. A few studies have also shown that GWs with wavelengths larger than 1000 km are rare in thermosphere during geomagnetically quiet times \cite{Garcia16}. The medium-scale GWs can either be primary waves that penetrate directly from the lower atmosphere or higher-order GWs \cite{Vadas13, Becker18}.

This article is organized as follows. Section 2 describes the WAM model, simulations and the methodology used to extract GWs from the model output. Section 3 presents the results and discusses the GW climatologies derived from WAM as a function of season, altitude, local time, and longitude. Section 4 contains the summary and some concluding remarks. 

\section{Methodology}
\subsection{Model}
The Whole Atmosphere Model (WAM) is an extension of the Global Forecast System (GFS) model by National Weather Service \cite{Akmaev08, Akmaev08b, Akmaev11}. The GFS model was developed to provide accurate medium-scale tropospheric weather forecasts. The combination of WAM and the Ionosphere-Plasmasphere-Electrodynamics (IPE) model has been used by the National Oceanic Atmospheric Administration operations to provide short-range space weather forecasts since July 2021. The WAM spans the Earth's atmosphere from the surface to geometric height of about $\sim$600 km (depending on solar activity). It is a hydrostatic model with a spectral dynamical core \cite{Sela09}. The current operational version of WAM, WAMT62 is based on the GFSv13 atmosphere model and uses a triangular spectral truncation at a total horizontal wavenumber of 62, which is equivalent to horizontal grid spacing of $\sim$180 km. This model has 150 levels in the vertical, with a vertical resolution of $1/4$ pressure scale height. 

The WAM includes all the GFS physical parameterizations for lower atmospheric physics, such as radiative transfer, hydrological cycle, planetary boundary layer and surface exchange processes, orographic and non-orographic GW schemes, and vertical eddy diffusion. It uses the lower tropospheric orographic gravity wave drag parameterization as described in \citeA{Kim95} with mountain blocking effects according to \citeA{Lott97} and \citeA{Kim05}. The model subgrid scale orography is described as in \citeA{Baines90} and \citeA{Lott97} and is a function of model resolution. The IT physics includes the parameterizations for UV and EUV radiation, non-LTE infrared radiative cooling (O$_3$, CO$_2$, and H$_2$O), NO cooling, ion drag, Joule heating, molecular diffusion, and conductivity. The high-latitude electric potentials and magnetic field-aligned currents are represented by \citeA{Weimer05}, and particle precipitation patterns by \citeA{Fuller87}. 

The high-resolution version of the model, WAMT254 extends the spectral domain up to wavenumber 254 and has been developed to study smaller-scale processes in the IT region. It has a horizontal resolution of $\sim$52 km and thus can effectively resolve waves of horizontal wavelengths of $\sim$200 km and above. It has 150 full model layers with similar vertical resolution as the T62 version. The whole atmosphere modeling regime enables us to investigate the propagation of medium-scale waves from lower to upper atmosphere without introducing any boundary effects and invoking GW empirical parameterizations. The simulations in this study are performed for the full months of March, June, September and December with fixed idealized solar wind and geomagnetic conditions, with Kp=1 and F10.7=70. Using geomagnetic quiet conditions eliminates conditions that can generate Traveling Atmospheric Disturbances (TADs) at high latitudes that can overpower quiet-time thermospheric GWs from the lower atmosphere.

\subsection{Zonal Mean Wind Climatology} 
\begin{figure}
\includegraphics[width=\textwidth]{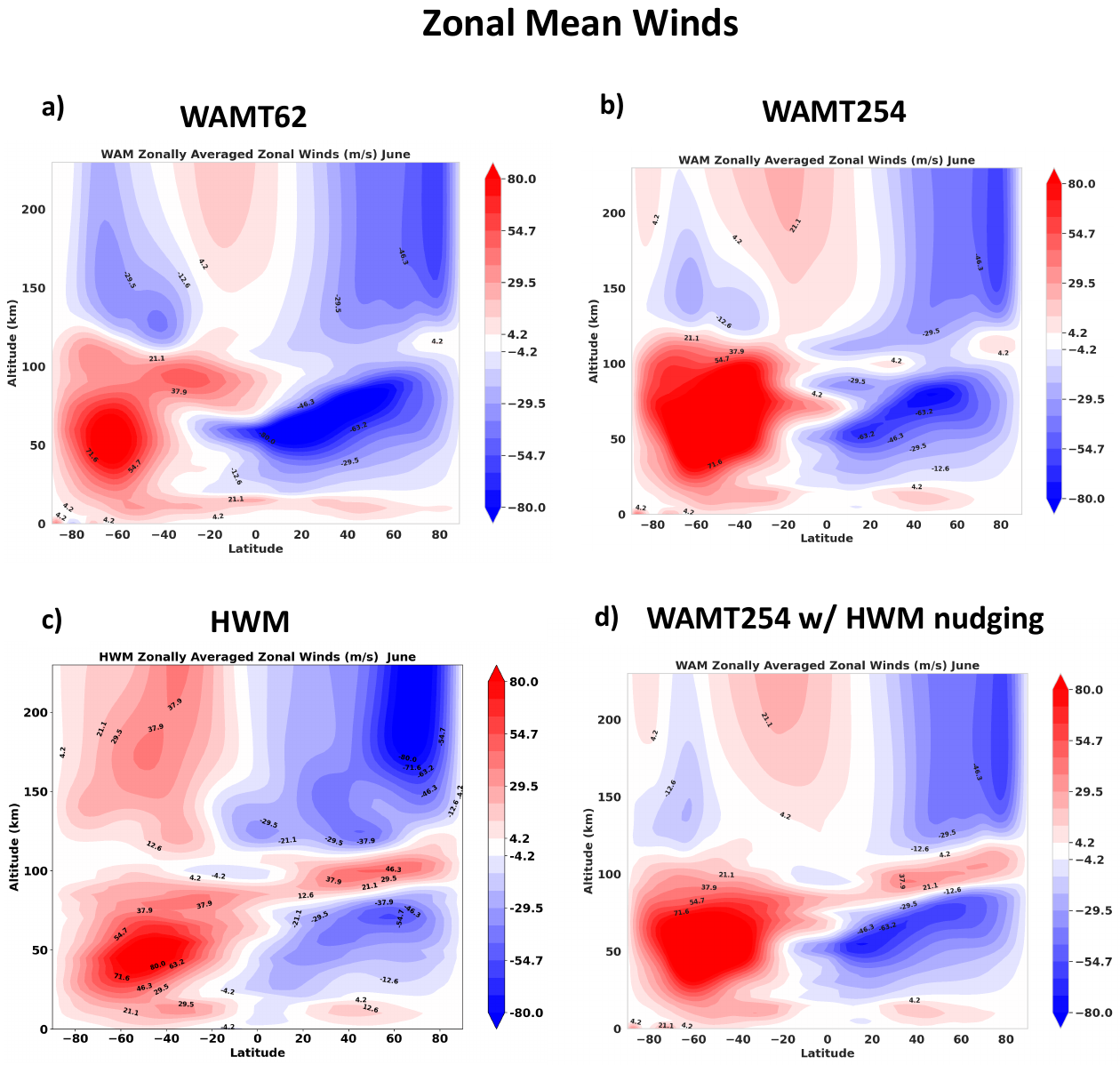}
\caption{Zonal winds (m/s) averaged zonally and over the month of June for a) WAMT62, b) WAMT254, c) HWM empirical model, d) WAMT254 with HWM nudging in the MLT region.}
\label{merid_winds}
\end{figure}

Figure \ref{merid_winds} shows the zonal-mean zonal wind for June from WAMT62 (panel a) and WAMT254 (panel b) compared with the empirical Horizontal Wind Model (HWM14, panel c) \cite{Drob15}. The winds in the HWM are derived from a wide variety of sources, e.g., extensive sounding rockets, radiosondes database in the lower atmosphere, and incoherent scatter radar, optical satellite measurement in the IT region \cite{Drob08}. In the MLT region, the winds from the UARS High Resolution Doppler Imager (HRDI) and Wind Imaging Interferometer (WINDII) instruments are included, which have been extensively studied \cite<e.g.,>{Burrage96, Lieberman93, McLandress96}. The wind fields from WAMT62 and WAMT254 show climatological features similar to HWM below $\sim$90 km. The winter (Southern) hemisphere is dominated by the eastward winds, while summer (Northern) hemisphere has westward winds in the stratosphere up to the upper mesosphere.

At around 100 km the HWM shows the expected reversals of extratropical zonal winds, with strong eastward (weak westward) winds in the summer (winter) hemisphere. The wind reversal in the summer hemisphere is caused by gravity wave momentum deposition, while the weak reversal at subtropical latitudes in the winter mesopause region is not well understood. Overall, GWs are responsible for the summer to winter circulation and the cold summer mesopause \cite{Lindzen81, Garcia85, Holton83}. In the summer MLT, WAMT62 does not show a wind reversal, whereas WAMT254 shows a slight improvement with weak eastward winds. In the winter MLT, both WAMT62 and WAMT254 show strong eastward winds. \citeA{Becker18} found that this feature is caused by momentum deposition of eastward propagating secondary GWs that are generated in the stratopause region and has been previously observed in other GW-resolving GCMs as well \cite<e.g.>{Miyoshi08, Miyoshi14, Watanabe08, Watanabe09, Sato12, Liu17}. In WAMT62, the GWs that contribute to the momentum budget in the MLT region are under-resolved, and hence it would require GW parameterization for realistic zonal winds \cite<e.g.>{Yudin20}. However, the lack of strong wind reversal in the summer MLT in WAMT254 can be because of several reasons. 

One is that the T254 spectral resolution is not enough to resolve the important spectrum of GWs that maintains the MLT summer to winter circulation. \citeA {Preusse08} showed that waves with wavelengths smaller than $\sim$150 km can also propagate to the mesopause height and contribute to GW drag in the MLT. \citeA{Liu14c} showed that resolved GW drag caused the MLT jet reversals to be at slightly higher altitudes in the high-resolution WACCM-X (0.25$^o$ horizontal resolution), and was found to be insufficient to reverse the MLT jet. However, previous studies by \citeA{Watanabe09, Becker18} have achieved near-realistic MLT winds in their global models with a horizontal spectral resolutions of T212 and T240, respectively, which is coarser than WAMT254. Thus, there is currently a lack of understanding regarding which part of the GW spectrum transports what amount of horizontal momentum deposition into the MLT. It is also likely that the current hyperdiffusion applied in WAM is too dissipative for medium-scale waves. \citeA{Becker09b} discussed that employing a physics-based non-linear macro-turbulent diffusion scheme in the lower and middle atmosphere that is based on the ideas of \citeA{Smago63, Smago93} is able to better handle the energy cascades that lead to interactions between the GWs and the mean flow in GCMs. Efforts are underway to improve the WAM model accordingly. 

In the middle-upper thermosphere, in the winter hemisphere, HWM has eastward winds whereas both WAMT62 and WAMT254 have westward winds at high latitudes and eastward at mid-low latitudes. Since WAM simulations are run under geomagnetically quiet conditions, westward winds at high latitudes are less likely due to high-latitude ion drag, and more likely a result of bias in the model. Since wind measurements in this region are few and sparse, more thermospheric measurements are required to understand the correct global thermospheric wind distribution.

Gravity waves interact with the background winds and are filtered as they propagate from the lower atmosphere to the IT region. Since the goal of this study is to investigate the quiet time thermospheric GWs, which result either directly from the lower atmosphere or are secondary and higher-order GWs resulting from multi-step vertical coupling \cite<e.g.>{Vadas19}, it is important to have more realistic background winds in WAM MLT. The above-mentioned shortcomings of WAMT254 in the zonal mean climatology can also be addressed by incorporating GW parameterizations. However, in our diagnostic study of resolved GWs, we decided to improve the zonal mean winds without using GW parameterizations. We use a coefficient-based nudging technique in grid-space to nudge the WAM MLT winds towards HWM. We only nudge the background zonal and meridional winds in the MLT, that is the harmonics with small wavenumbers, without altering the gravity wave spectrum. Figure \ref{merid_winds}d shows that the zonal mean zonal winds in WAMT254 after HWM nudging agree better with the HWM winds in the MLT region. We use the nudged-WAMT254 model for the simulations discussed in this study, and simply refer to it as WAM.

\subsection{Extracting Gravity Waves}
\begin{figure}
\includegraphics[width=\textwidth]{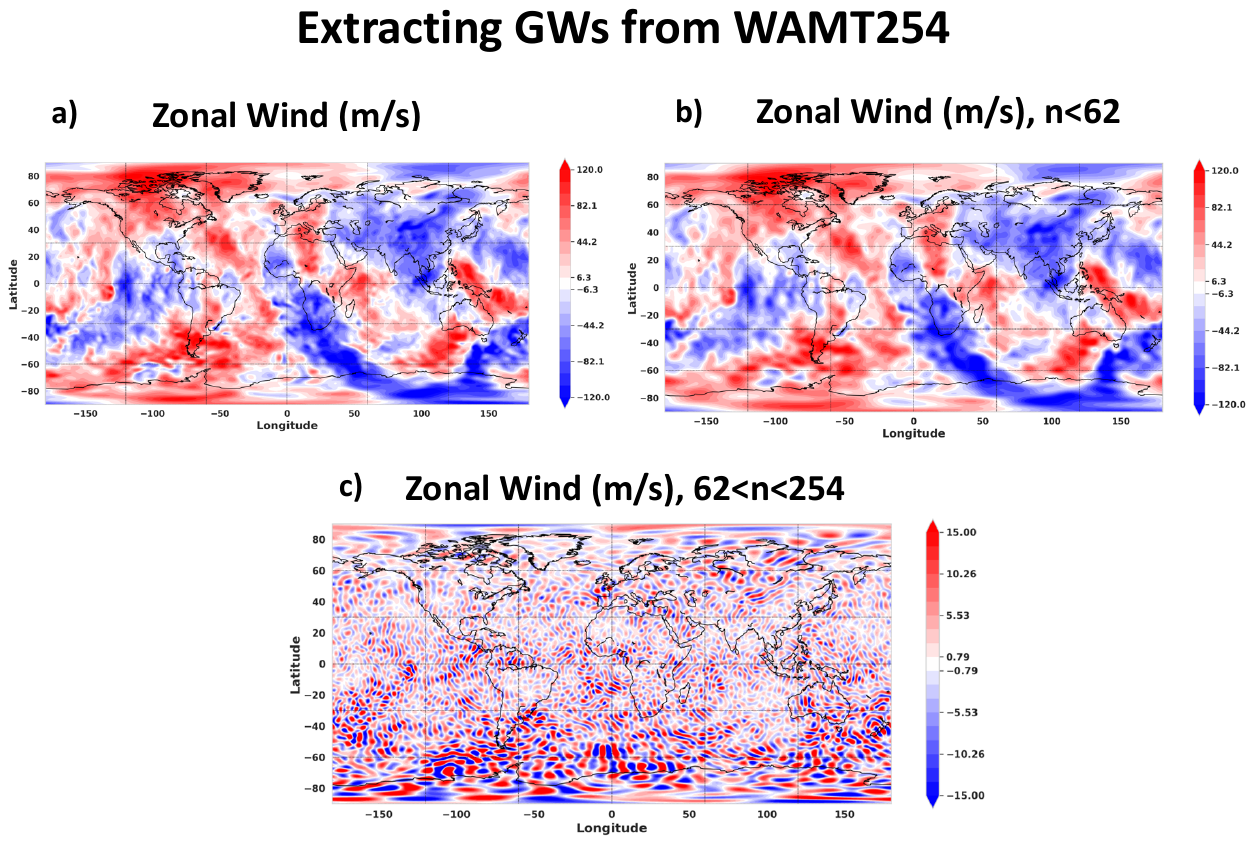}
\caption{a) Example of WAM Zonal Winds (m/s). b) Truncated Zonal Winds (m/s) with wavenumbers$<$64, showing only the larger-scale waves. c) Difference between panels a) and b), showing the zonal winds with 64$<$wavenumbers$>$254, thus demonstrating the GW activity (m/s).}
\label{filter}
\end{figure}

We extract gravity waves from the output of the nudged-WAMT254 model as a post-processing step. Figure \ref{filter}a shows an example of the zonal wind output from the model at a specific time step and a vertical level. Figure \ref{filter}b shows the zonal winds with only the large-scale waves, obtained after truncating the harmonics of the original output (Figure \ref{filter}a) till total horizontal wavenumber 62. The truncated zonal winds are then subtracted from the original model output to extract wavenumbers$>$ 62, which is shown in Figure \ref{filter}c. This method extracts all the waves with scale sizes from $\sim$200 km (smallest that is resolved by the model, i.e., wavenumber 254) till about $\sim$600 km (wavenumber 62). Our results in this study comprise largely of GW activity derived from zonal winds. This allows us to leverage previous observations of GWs in zonal winds in the thermosphere from the CHAMP and GOCE accelerometer data for validation \cite<e.g.,>{Liu17, Park14, Park23}. The parameter used to describe GW activity is the square of the wave perturbation due to GWs, u'$^2$, also referred to as the GW variance. We use this parameter because it is proportional to the GW kinetic energy \cite{Tsuda00, Hei08} and can be easily determined from the above-mentioned in-situ thermospheric wind observations. Moreover, assuming linear theory of GWs, the ratio between the kinetic and potential energy of medium-to-high frequency GWs is 1 \cite{Tsuda00}, thus implying that the global distribution of GW kinetic and potential energy is roughly the same. Only when the intrinsic frequency is small (close to a breaking/critical level), kinetic energy becomes larger than potential energy.

\section{Results and Discussion}

\subsection{Lower Atmospheric Gravity Waves}

\subsubsection{Tropospheric Gravity Waves}
\begin{figure}
\includegraphics[width=\textwidth]{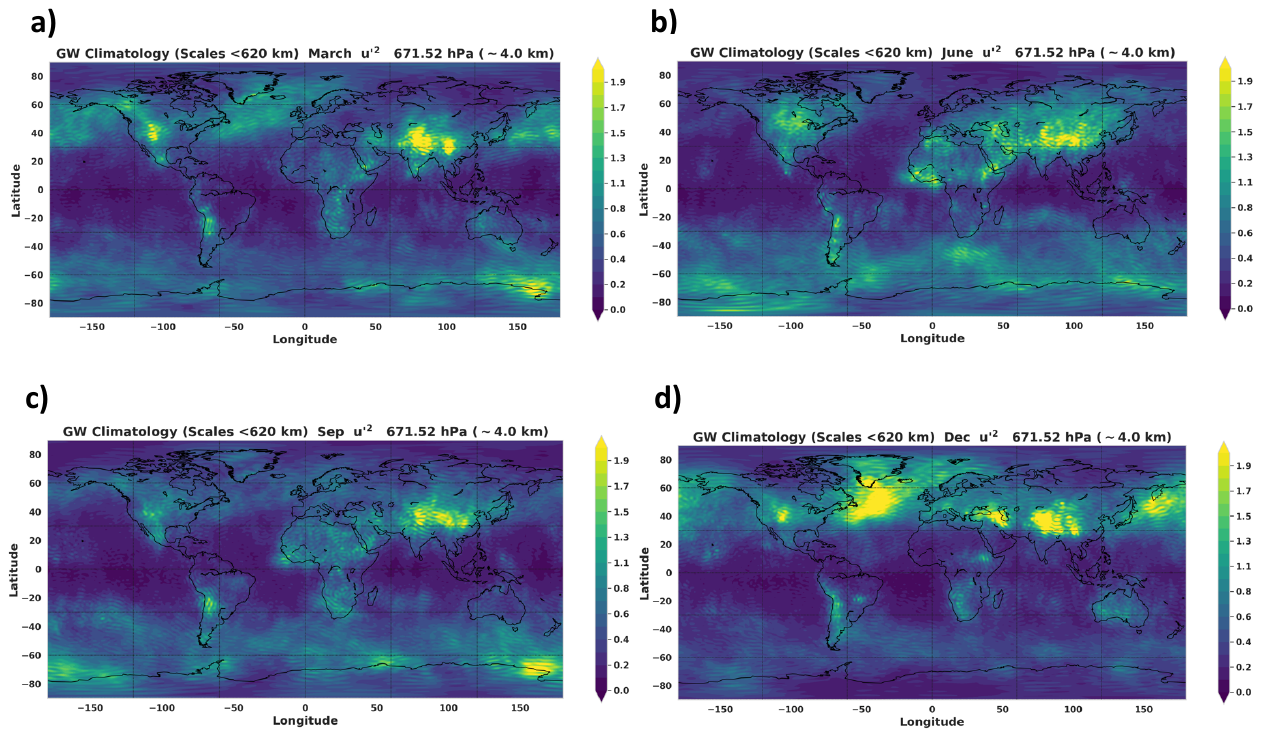}
\caption{WAM zonal wind variances (m$^2$/s$^2$; $\lambda$\textless620 km) at $\sim$4 km altitude averaged over a month for a) March, b) June c), September and d) December.}
\label{map_0}
\end{figure}

Figure \ref{map_0} shows the monthly averaged GW activity in the troposphere, at $\sim$4 km in WAM for different seasons. GW hotspots include North America, Eurasia and the mountain ranges such as the Andes, Himalayas. GWs are generated near the surface by the flow of air over mountains. A large activity is also observed around the northern Atlantic Ocean, northern Pacific, and Antarctica. During December, the southern hemisphere GW activity becomes more prominent over the Andes, southern Africa and Australia. Similar hotspots have been simulated for different resolutions of the European Centre for Medium-Range Weather Forecasts (ECMWF) IFS model and ERA reanalyses data \cite{Preusse14, Li20, Wei22}. Apart from the orography, non-orographic sources, e.g. convection, shear instabilities, also contribute significantly to GW activity. The next section discusses the GW activity in the stratosphere which includes GWs from these different sources.

\subsubsection{Stratospheric Gravity Waves}
Figure \ref{map_40} shows the comparison between the global distribution of WAM GW activity in the stratosphere (at $\sim$40 km) and the HIRDLS observations obtained from the gravity wave climatology based on atmospheric infrared limb emissions (GRACILE) dataset \cite{Ern18}, for different seasons. GRACILE is a dataset of satellite observations of GW global climatology in the stratosphere and mesosphere by \citeA{Ern18} and is based on the space-borne vertical temperature profiles from the infrared limb sounding satellite instruments, High Resolution Dynamics Limb Sounder (HIRDLS) onboard the Aura satellite \cite{Gille08} and Sounding of the Atmosphere using Broadband Emission Radiometry (SABER) onboard the TIMED satellite \cite{Yee03}. We only show the HIRDLS observations here because it has a horizontal sampling step of $\sim$90 km, and thus is able to observe the GWs of wavelengths smaller than 600 km, which are the focus of this study. SABER on the other hand has a sampling step ranging between 180-300 km, depending on the altitude. HIRDLS also has a continuous coverage of latitudes between 63$^o$S-80$^o$N, whereas SABER switches between the northward and southward viewing modes every $\sim$60 days, providing continuous coverage only of latitudes between 50$^o$S-50$^o$N. The HIRDLS GW variances are derived from the absolute momentum flux of GWs by using values of vertical and horizontal wavenumber provided by GRACILE.

Overall, the GW variances observed in HIRDLS temperature data are much larger than those resolved by the WAM dynamics. This is because in deriving GW activity from WAM we only include the wavenumbers$>$64, whereas the absolute momentum flux in GRACILE was estimated from the vertical temperature profiles employing the temperature background as a superposition of the zonal mean field and the first six zonal wavenumbers \cite{Ern18}. However, assuming similar sources for larger and smaller-scale GWs in the stratosphere, we expect the global distribution to be qualitatively similar for both. In this altitude regime, GWs spread out over a wider spatial area as compared to the tropospheric activity of Figure \ref{map_0}. The GWs smear out forming enhanced latitudinal band-like signatures and concentrated GW hotspots become less pronounced, signifying their horizontal propagation with altitude \cite{Trinh17}. During June and September, significant GW activity eastward of Andes is observed in both WAM and the HIRDLS data. The region above the Southern Andes and Antarctic Peninsula is a prominent source of mountain waves between the months of April to October \cite{Eckermann99, Wang10, Ern18}. The north-south orientation of the Andes acts as a long mountain barrier for favorable generation of orographic GWs when winter winds are strongly eastward \cite{Yan10}. Non-orographic GWs are also observed during local winter in both the hemispheres at high latitudes, as a result of various sources such as jets and fronts in the troposphere \cite<e.g.>{Hendricks14, Bossert20} and the polar vortex \cite<e.g.>{Vadas23}.

At low-mid latitudes during the warm season (NH in June and SH in December), GW activity is largely due to moist convection. Since neither convection nor the typical scales of convectively generated GWs are resolved in WAM and the resolved GWs generated by parameterized convection might have some unrealistic wave characteristics, the comparison of Figure \ref{map_40} shows a greater difference between WAM and the HIRDLS data in these regions. For e.g., In HIRDLS data, during June and at around 20$^o$N, large activity is observed over the Gulf of Mexico, North Africa, and over Southeast Asia. Whereas, during December, at 20$^o$S, GW sources are observed over South America, South Africa, Australia, and the Southern Pacific Ocean. These characteristic sources have also been previously observed by \citeA{Ern04, Preusse09, Yan10}. In Figure 5, these distinctive sources of convective GWs over the tropics are only observed during June in WAM and are not as clear during the other months because of the aforementioned limitation of whole-atmosphere models regarding convectively generated GWs. Moreover, in HIRDLS, this convective GW activity is around 20$^o$ off the equator in the respective summer hemispheres, whereas in WAM, it is largely present around the equatorial belt. It is likely that the GW spectrum simulated in WAM as a result of parameterized convection does not include the larger-scale waves of the GRACILE analyses and therefore the tropical GWs in WAM do not propagate meridionally as far as the corresponding waves in the GRACILE analyses, thus resulting in maxima at different latitudes. These differences will be the subject of future investigations in WAM.   

During September, in WAM and HIRDLS, dominant GW hotspot is eastward of Andes. During both March and September, weak convective activity is observed around the tropics. One of the major differences between WAM and HIRDLS is the large activity over the Tibetan plateau and Asia during March and December in WAM. Because of the large-scale mountains, Tibetan plateau has been previously observed and modeled as one of the major sources of orographic GWs \cite{Eckermann99, Cohen16, Li20, Wei22}. The HIRDLS climatology in GRACILE only contains GWs with vertical wavelengths$<$25 km \cite{Ern18} and hence can potentially underestimate GW variances in this region. Overall, we find that the global distribution of WAM GW activity has some differences with the HIRDLS observations but qualitatively captures the major global GW hotspots. In the next section, we will discuss WAM GWs in the middle thermosphere and compare it with those derived from the in-situ measurements of GOCE satellite. 

\subsection{Thermospheric Gravity Waves}
\subsubsection{Seasonal Variations}
 
Figure \ref{map250} shows the comparison between GW variances from WAM and Gravity Field and Steady-State Ocean Circulation Explorer satellite (GOCE) in the thermosphere at an altitude of $\sim$250 km. GOCE is in a sun-synchronous orbit and covers only the dawn-dusk local time sectors. For better comparison, WAM GW activity shown in this figure is averaged over the dawn and dusk local times for each season. The accelerometers onboard GOCE are used to derive in-situ cross-track neutral winds, which are then processed to derive zonal and meridional winds \cite{Doornbos13, March19}. To extract GWs from GOCE winds, we use the technique similar to \citeA{Park14} and \citeA{Liu17}. To summarize, we apply a low-pass filter along the satellite track to derive mean background zonal winds (u$_o$), which are then subtracted from the total zonal winds (u) to determine GW perturbations in zonal winds (u'). In accordance with the orbital speed of GOCE and sampling cadence of 10s, we use a low-pass filter of window size 11 to extract waves of scale sizes between 200-600 km. The GW variances for 2010-2013 are then binned into latitude and longitude grids and averaged over solstices and equinoxes. We also filtered GOCE data by geomagnetic activity such that we only include data when Kp$\leq$1.

\begin{figure}
\includegraphics[width=\textwidth]{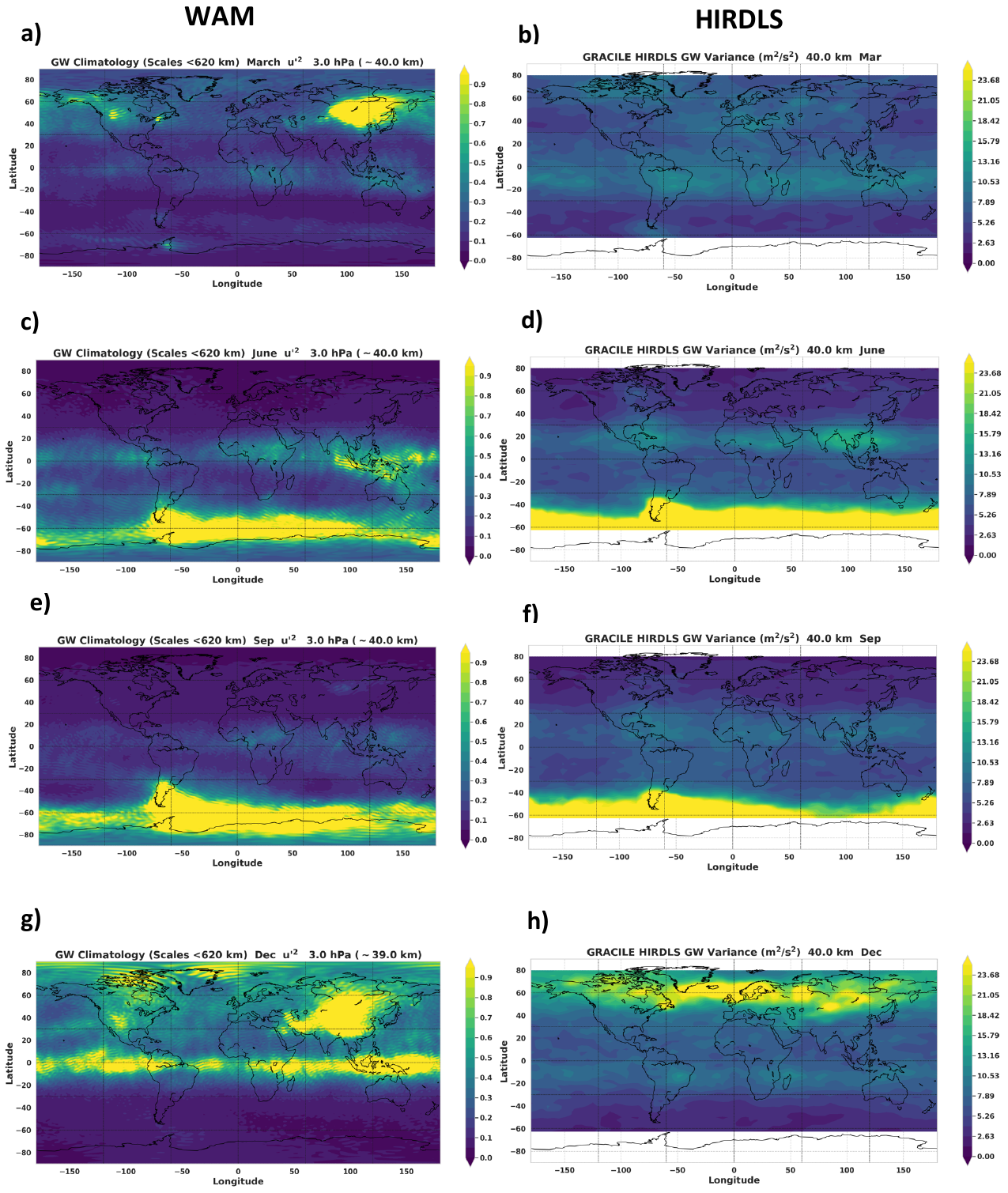}
\caption{Monthly averaged WAM zonal wind variances (in m$^2$/s$^2$; $\lambda$\textless620 km) compared with the wind variances observed in the HIRDLS data (from the GRACILE empirical model), at $\sim$40 km altitude. a) and b) show this comparison for March, c) and d) for June, e) and f) for September, and g) and h) for December.}
\label{map_40}
\end{figure}

\begin{figure}
\includegraphics[width=\textwidth]{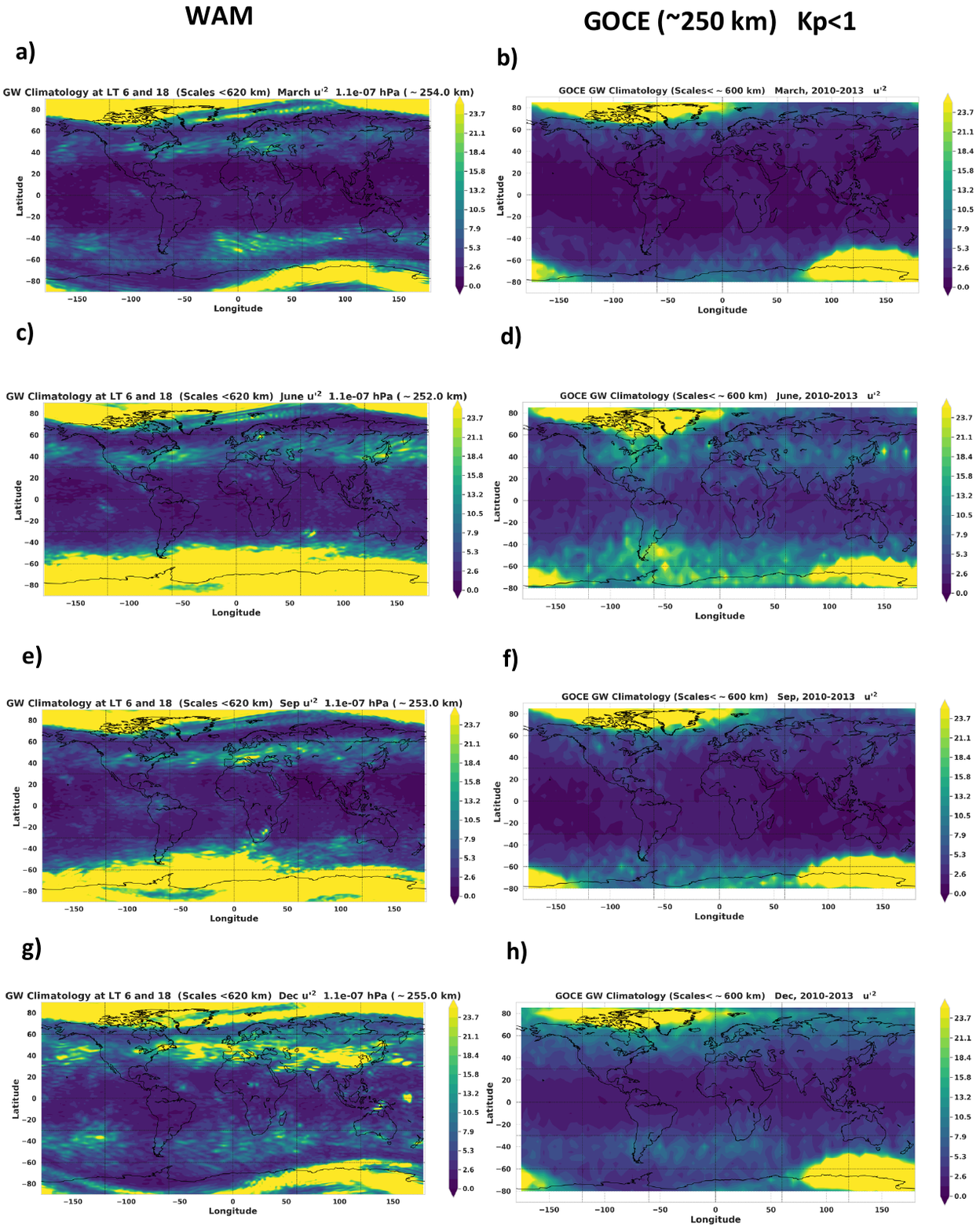}
\caption{Monthly averaged WAM zonal wind variances (m$^2$/s$^2$; $\lambda$\textless620 km) at an altitude of $\sim$250 km compared with the zonal wind variances observed in the GOCE accelerometer data. WAM variances are averaged over only 6 and 18 LT to replicate the dawn-dusk orbit of GOCE satellite. GOCE data is only taken for days when Kp$\leq$1. a) and b) show this comparison for March, c) and d) for June, e) and f) for September, and g) and h) for December.}
\label{map250}
\end{figure}

In the thermosphere, GW activity spreads globally \cite<e.g.,>{Trinh17, Malhotra21} and the magnitudes of variances are much larger than in the troposphere and stratosphere. Even though we do not distinguish between primary and higher-order waves in WAM analyses, we note that GWs reaching this altitude are very likely a combination of primary and higher-order waves \cite{Vadas14, Forbes16}. The GW activity is most prominent at high latitudes in all seasons in both the model and GOCE observations. During June and September, secondary and higher-order GWs of orographic origin from the Andes mountains contribute to the variances in the southern hemisphere (also observed in the stratosphere in Figure \ref{map_40}). Mountain waves from the Andes can penetrate into the middle atmosphere during local winter because the strong background winds can preferentially refract GWs with low phase speeds to longer vertical wavelengths \cite{Jiang02, Preusse08}. These waves can then generate higher-order waves which likely reach the IT heights as shown in this figure \cite{Vadas19, Becker20}. The magnitudes of GW activity are slightly higher in WAM than in GOCE data, especially in this region during June and September. 

In both WAM and GOCE, during solstices, the GW activity has hemispherical asymmetry with larger activity in the winter than in the summer hemisphere between 40$^o$-90$^o$ latitudes. A large proportion of the mid-high latitude winter activity is due to GWs with their sources in the stratosphere (Figure \ref{map_40}) \cite{Sato00, Li09, John12, Becker22b, Vadas23}. A band of GW activity is also observed in the summer mid-latitudes between 40$^o$-60$^o$ latitudes in June and December in both the model and data. These waves are likely higher-order waves originating from the low-frequency waves in the tropical belt that are generated by convection (as shown in the stratosphere in Figure \ref{map_40}), and that move poleward with altitude \cite{Preusse09, Ern11, Trinh17}. Similar mid-latitude GW activity is also seen during equinoxes in WAM but not in GOCE data. 

At low-latitudes, the GW activity is weak in both WAM and GOCE. During March, GOCE shows a distinguishing signature of slightly larger GW activity over land as compared to oceans. However, this longitudinal dependence is not observed in WAM and can be quite sensitive to the convective GW scheme used in the model. Using satellite observations, this feature has been previously observed by \citeA{Liu17} in the same dataset, \citeA{Park14} in CHAMP and \citeA{Park23} in GRACE data. These are likely higher-order waves originating from mountain lee waves that are generated by flow over topography \cite{Tsuda00}. It is not as prominent in GOCE analyses shown here because of different averaging windows. All of the above-mentioned studies have temporal averages over longer periods of time of $\sim$3 months in contrast with the 1-month averages shown here. Longer-term averages for high-resolution GCM outputs present a significant computation challenge and hence are not compared here. 

During all the seasons, large activity is observed around the geomagnetic poles in both the model and data. Previous studies have discussed that small-scale GWs concentrated over the geomagnetic poles can be generated due to different sources such as the Lorentz force of the auroral electric currents, Joule heating, and energetic particle precipitation \cite{Forbes16, Trinh17}. However, we believe that in-situ generation of GWs is unlikely to be a major source in the WAM simulations shown here as we have used idealized quiet geomagnetic conditions. The source of these waves remain controversial. Some previous studies have suggested that these GWs are of lower atmospheric origin, and are focused towards higher latitudes with increasing altitude, because of oblique propagation of GWs and their interaction with stronger high-latitude background winds \cite{Sato09, Miyoshi14, Garcia16}. However, recent studies by \citeA{Vadas19} and \citeA{Becker20} suggest a strong influence of higher-order GWs at high-latitudes.

Overall, we find that the magnitudes and global distribution of thermospheric GW activity in WAM is in reasonable agreement with GOCE observations and previous studies. Possible reason for higher magnitudes of GW activity in WAM relative to GOCE can be related to the numerical framework of the model and the diffusion scheme used. Apart from seasonal variations, only few studies have found a significant variation of thermospheric GW activity with local time \cite<e.g.,>{Miyoshi14, Park23}. In the IT region during geomagnetically quiet times, since tidal pressure gradients dominate the mean background wind field, GWs are heavily influenced by tides \cite<e.g.,>{Miyoshi08, Becker22}. Specifically, the in-situ generated migrating diurnal tide can potentially introduce a local time dependence in thermospheric GW activity at middle and high-latitudes. Considering reasonable representation of thermospheric GWs in WAM, their local time dependence is investigated in the next section.

\subsubsection{Local Time Dependence}
\begin{figure}
\includegraphics[width=\textwidth]{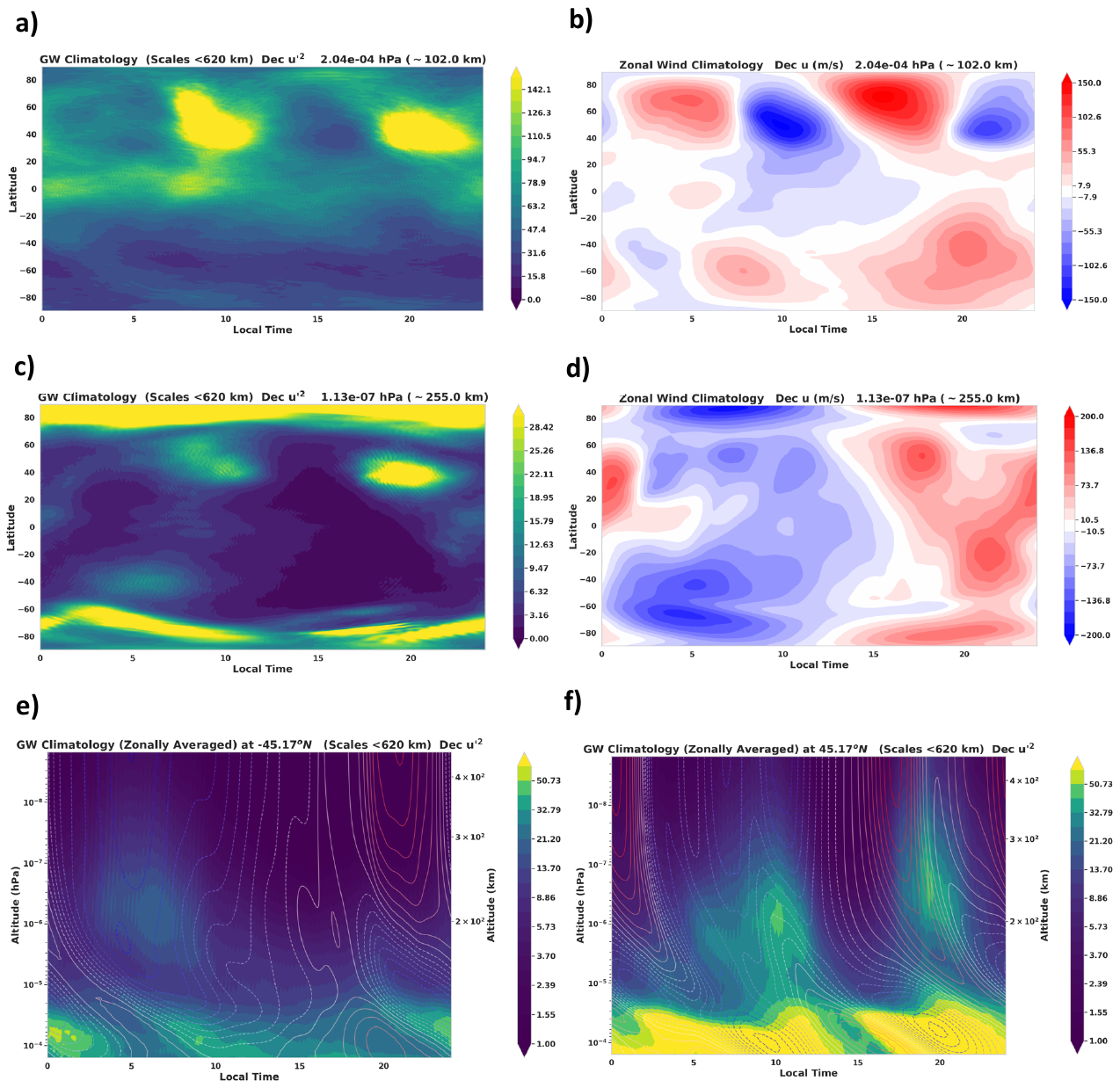}
\caption{a) and b) Monthly and zonally averaged WAM zonal wind variances (m$^2$/s$^2$; $\lambda$\textless620 km) compared with the zonal mean winds at $\sim$102 km, plotted as a function of latitude and local time. c) and d) show the same comparison for $\sim$255 km, e) and f) show zonally averaged zonal wind variances plotted as a function of altitude and local time for December at $\sim$45$^o$S and $\sim$45$^o$N, respectively.}
\label{lt_lat_low}
\end{figure}

As GWs propagate upward from the lower atmosphere, they interact with the mean background winds \cite<e.g.,>{Vadas05, Yamashita13, Miyoshi14}. Since, the background zonal wind does not show a strong local time dependence below $\sim$80 km \cite<e.g.>[Figure 4]{Miyoshi14}, they do not exhibit significant local time dependence of GW activity in the lower atmosphere. Hence, we only discuss the local time dependence of GW activity and its relation to the background zonal winds in the thermosphere. Panels a) and c) of Figure \ref{lt_lat_low} show the zonally averaged thermospheric GW variances at $\sim$102 km and $\sim$255 km averaged for the month of December, and its dependence on latitude and local time. Panels b) and d) show the same for zonal mean zonal winds. The zonal winds of WAM have been previously found to be in reasonable agreement with CHAMP winds during March and December by \citeA{Lieberman13} (their Figure 1). Averaging zonally for each local time eliminates the effect of non-migrating tides, while retaining only the background mean winds and the migrating tides
 
In the mesopause region, at $\sim$102 km, the background winds show a strong semidiurnal variation at winter mid-high latitudes and predominantly eastward winds and weaker semidiurnal variation in the summer hemisphere (owing to the MLT wind reversal as shown in Figure \ref{merid_winds}). The GW activity shows a significant semidiurnal variation at winter mid-high latitudes and smaller magnitudes in the summer hemisphere. Comparing with the background zonal winds, it is evident that GW activity is larger when/where the background winds are westward in both the hemispheres, resulting in the semidiurnal variation in the winter hemisphere. In the winter mesopause region, this result is indicative of secondary GWs having eastward phase speeds \cite{Becker18}. Additionally, a similar dominance of eastward components of GWs in the thermosphere in winter as well as summer was also simulated by \citeA{Liu24} and \citeA[Figure 3]{Miyoshi14} using the Whole Atmosphere Community Climate Model with thermosphere/ionosphere extension (WACCM-X) and ground-to-topside model of atmosphere and ionosphere for aeronomy (GAIA), respectively. 

In the middle thermosphere, at $\sim$255 km, the background winds are dominated by strong diurnal variation at all latitudes with mostly westward winds between 0-12 LT and eastward winds after that \cite{Volland77}. Similarly, a diurnal variation is also observed in the GW activity at low-latitudes and in the summer hemisphere, such that a strong decrease in GW activity is observed after 12 LT at summer mid-high latitudes in the eastward background winds. However, in the winter hemisphere, the GW activity continues to show the semidiurnal variation. Previous studies by \citeA{Becker20, Becker22} have also observed strong local time dependence in meridional wind variances (not shown in this study). When the thermospheric tidal winds are poleward around noon, large-amplitude higher-order GWs were observed to propagate equatorward due to tidal-induced directional dissipation of GWs \cite{Fritts08b}.

Panels e) and f) show the altitude-local time dependence of GW variances for $\sim$45$^o$S and $\sim$45$^o$N respectively. The contours in both of these plots show the background mean winds, with dashed (solid) contours representing westward (eastward) winds. In the southern (summer) hemisphere, the GW activity in the MLT is weaker, owing to less GWs propagating up from the lower atmosphere (Figure \ref{map_40}). Weak semidiurnal variation in MLT GW activity is observed with maxima when background zonal winds are westward. The semidiurnal variation dissipates and transitions into largely diurnal variation above the altitude of $\sim$150 km with larger activity before 12LT. Since these GWs propagate to higher altitudes when the tidal zonal wind is westward before dissipating, they are likely to have predominantly eastward momentum flux. On the other hand, the GW activity reaching the northern (winter) hemisphere MLT is stronger due to selective filtering by background winds in the stratosphere and mesosphere \cite{Becker12}, while also exhibiting a strong semidiurnal component. GWs penetrate into the upper thermosphere between 5-12 LT when background thermospheric winds are strongly westward, and between 18-22 LT when background winds are westward in the MLT and weakly eastward above that. Thus, the strong semidiurnal variation of zonal wind variance in the winter MLT persists to higher altitudes, despite the dominant diurnal component in background zonal winds above $\sim$150 km. A deeper diagnoses of the GW characteristics, i.e. phase speeds, wavelengths, propagation direction is required to further understand these GW-tidal interactions in the thermosphere.

\begin{figure}
\includegraphics[width=\textwidth]{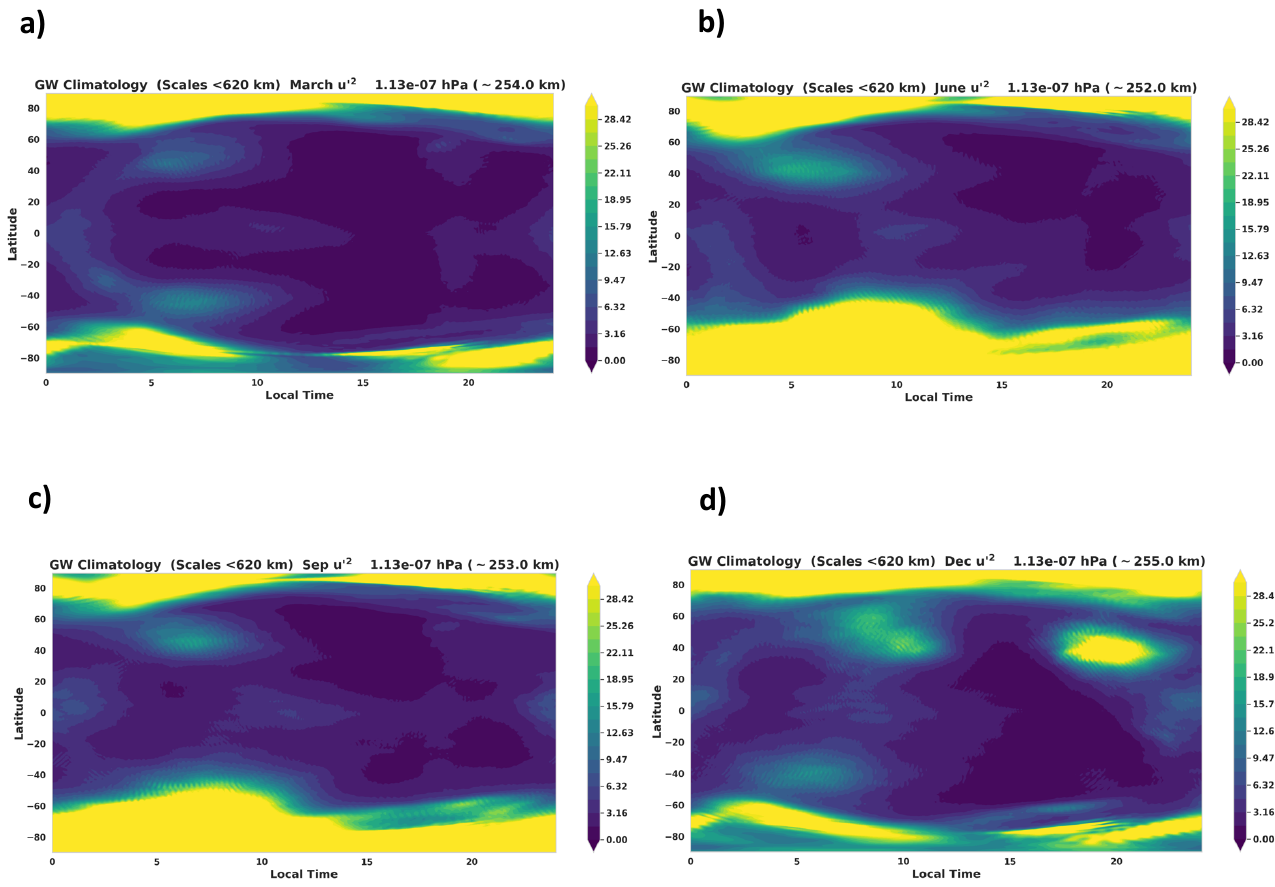}
\caption{Monthly and zonally averaged WAM zonal wind variances (m$^2$/s$^2$; $\lambda$\textless620 km) at $\sim$250 km, plotted as a function of latitude and local time for a) March, b) June, c) September, and d) December.}
\label{lt_lat}
\end{figure}

Figure \ref{lt_lat} demonstrates the local time dependence of thermospheric GW activity in terms of zonal wind variance for different seasons. During June and September, larger activity is observed in the Southern hemisphere due to orographic activity from the Andes as observed in the previous section. In all the seasons, a diurnal variation in GW distribution is seen at most latitudes, with larger activity during midnight-morning sector, agreeing with previous satellite observations \cite<e.g.>{Bruinsma08, Park23}. This diurnal variation is more prominent at solstices at low latitudes and high summer latitudes, while the winter hemisphere has a strong semidiurnal variation at middle latitudes. This results in an asymmetry with respect to season with minima in GW activity biased towards the summer hemisphere. \citeA{Park23} also reported a similar local time and seasonal asymmetry in GRACE(-FO) and CHAMP GW observations. In Figure \ref{lt_lat}, during equinoxes, similar to solstices, apart from the diurnal variation at most latitudes, semidiurnal variation is also observed at high-latitudes. In addition, a weak signature of terdiurnal component is also observed at low-latitudes. The high-latitude activity does not show a definitive local time dependence in all the seasons. This is because at high-latitudes, because of strong cross-polar flow of background winds, the GW activity has a stronger meridional component \cite{Forbes07, Becker22}, and meridional wind variance is not included in this study. 

Previous study by \citeA{Park23} have reasoned that the diurnal variation in GW activity can be due to the effect of kinematic viscosity on GW propagation. However, our results here demonstrate that migrating tides significantly affect the local time dependence of GW activity at thermospheric heights. In addition, the longitudinal dependence of GW activity is a relatively understudied topic. Similar to the migrating tides, it is likely that non-migrating tides also affect the propagation of GWs to higher altitudes in the thermosphere, thereby affecting longitudinal distribution of GW hotspots. This is discussed in the next section.

\subsubsection{Longitudinal Variations}
\begin{figure}
\includegraphics[width=\textwidth]{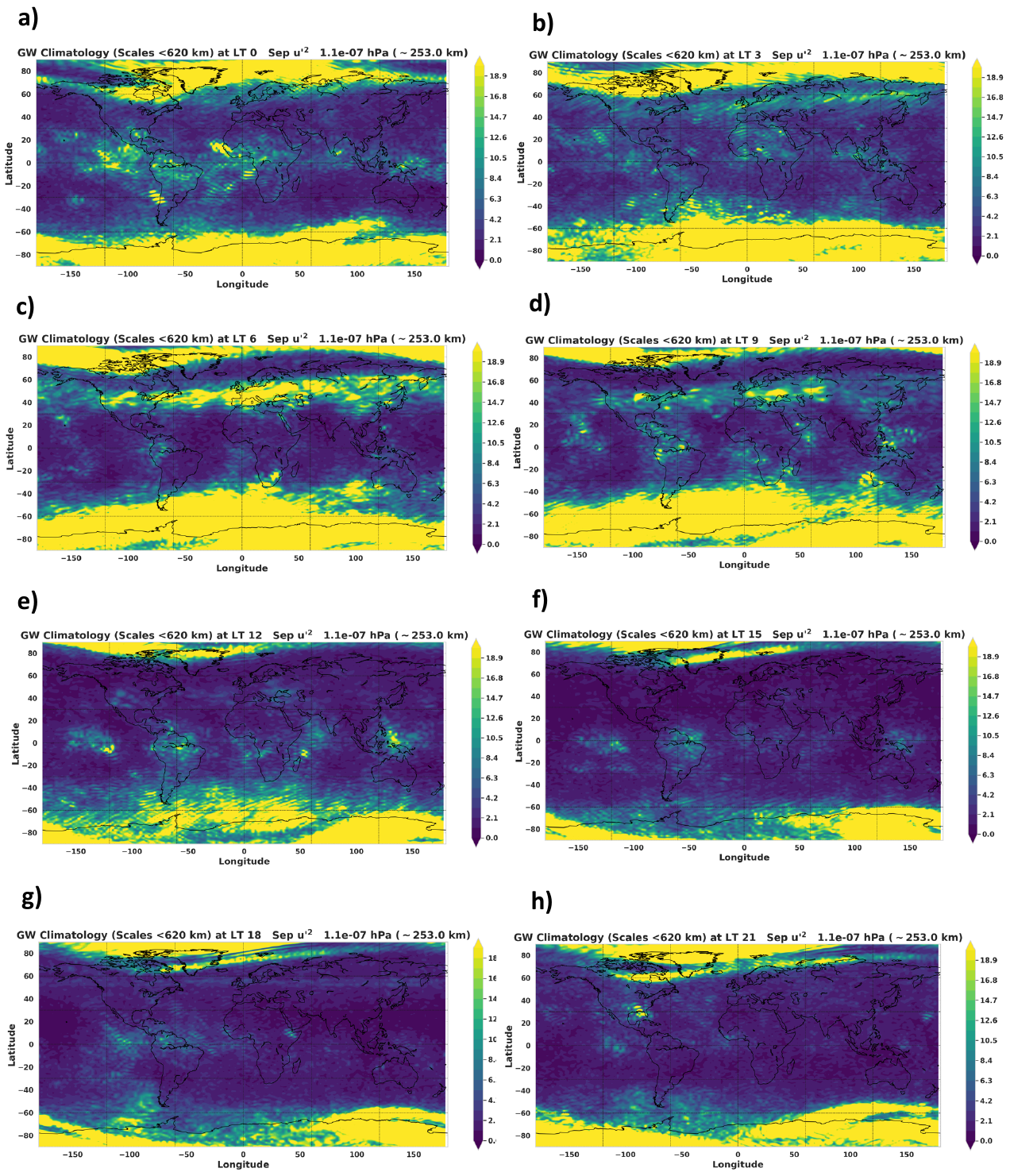}
\caption{WAM zonal wind variances (m$^2$/s$^2$; $\lambda$\textless620 km) averaged for September at an altitude of $\sim$250 km at fixed local times (LT), plotted as a function of latitude and longitude. Panel a) shows the distribution of GW activity at 0 LT , b) at 3 LT, c) at 6 LT, d) at 9 LT, e) at 12 LT, f) at 15 LT, g) at 18 LT, and h) at 21 LT.}
\label{lt_var}
\end{figure}

\begin{figure}
\includegraphics[width=\textwidth]{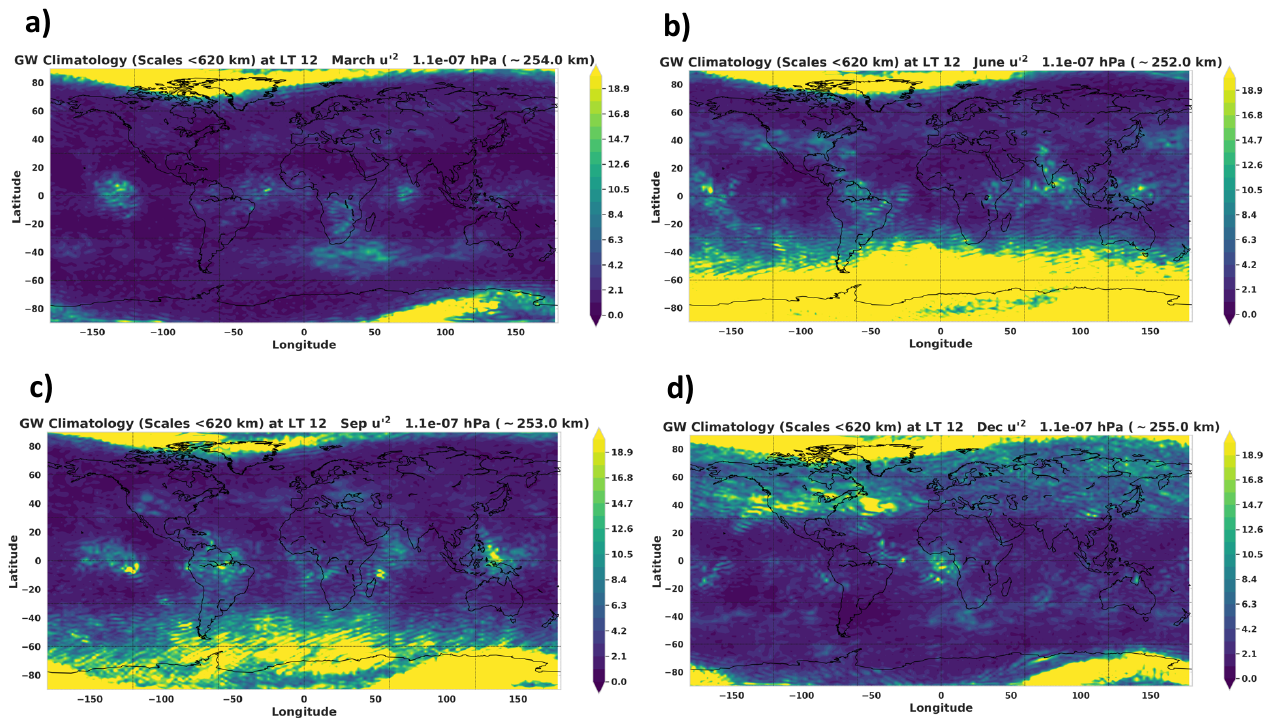}
\caption{Monthly averaged zonal wind variances (m$^2$/s$^2$; $\lambda$\textless620 km) at a fixed local time of 12 LT, and at an altitude of $\sim$250 km. Panel a) shows the latitude-longitude distribution of GW activity for March, b) shows for June, c) shows for September, d) shows for December.}
\label{seas_var}
\end{figure}

To understand the longitudinal dependence of GW activity, we change our coordinate system from UT-frame to LT-frame, i.e., from Earth-fixed to Sun-fixed coordinates. This helps in highlighting the influence of non-migrating tides in the GW activity. Figure \ref{lt_var} shows the spatial distribution of GW activity in terms of zonal wind variances at $\sim$250 km for September at different local times in 3-hour intervals. Each panel is averaged over one-hour local time. As discussed before, GW activity is larger between LT 0 and LT 12 at all latitudes. At high latitudes, the GW activity is similar to that discussed for Figure \ref{map250}. At low-latitudes, between LT 6 and LT 12, a longitudinal signature with wavenumber-4 is observed, which moves eastward with local time. The eastward motion is more pronounced in the video provided in the supporting material. This signature is most distinct at around noon-local time. Figure \ref{seas_var} shows the GW variance at LT 12 and at $\sim$250 km averaged for different seasons. The wavenumber-4 signature is the most prominent in June and September and almost disappears in December.  

\begin{figure}
\includegraphics[width=\textwidth]{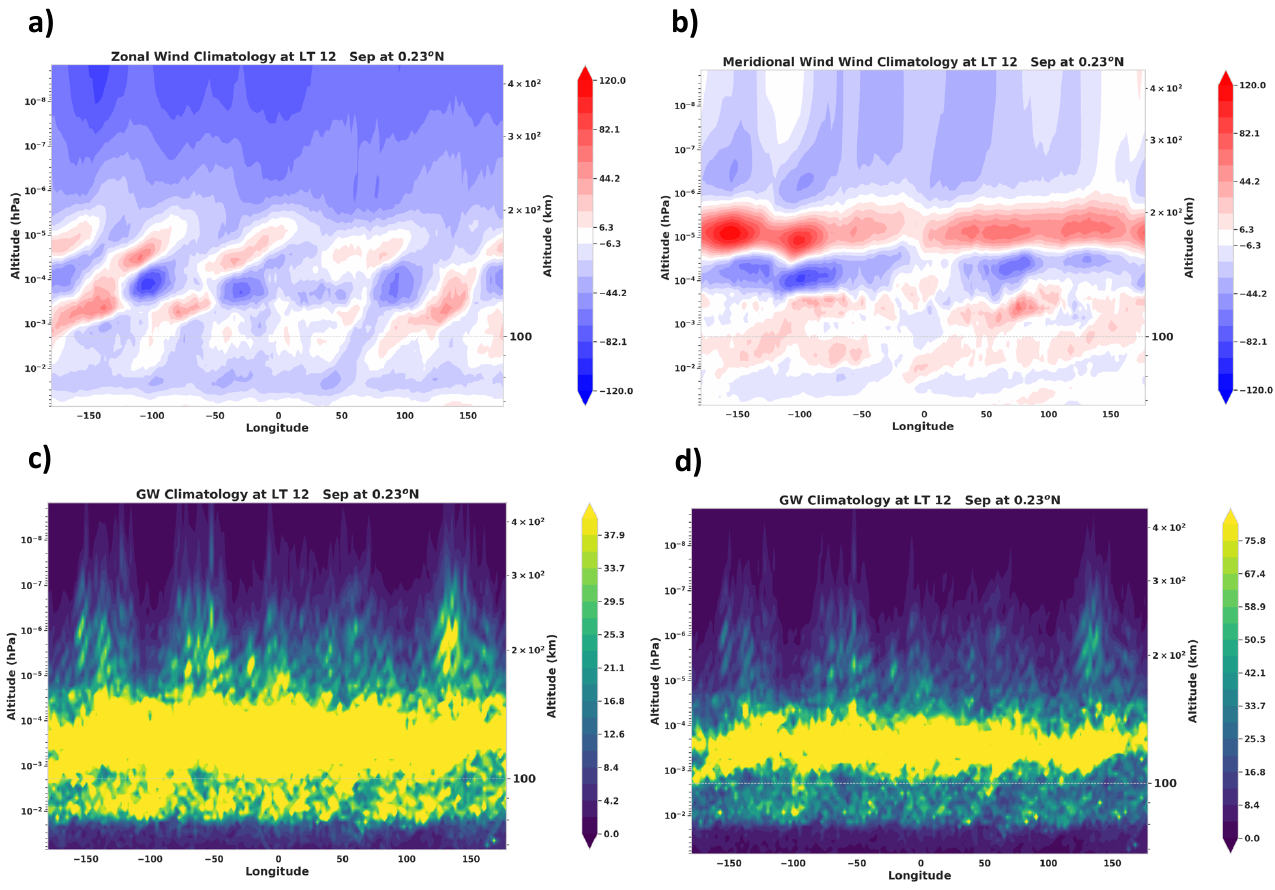}
\caption{Altitude-longitude distribution of monthly averaged thermospheric parameters for September, at 0.23$^o$N and at fixed local time of 12 LT. a) Zonal winds, b) Meridional winds. Panels c) and d) show GW variances in zonal winds  (m$^2$/s$^2$; $\lambda$\textless620 km) with different scales. These quantities are plotted against pressure in hPa on the y-axis, but the approximate altitude in km is also shown on the right y-axis.}
\label{lon_alt}
\end{figure}

Figure \ref{lon_alt} shows the altitude-longitude distribution of zonal winds (panel a) and meridional winds (panel b) at 12 LT at the equator for September. Panels c) and d) shows the similar distribution of GW activity (in terms of zonal wind variance) with different color scales. The background zonal winds show the strongest wavenumber-4 signature in the MLT region extending between 100-200 km and a weaker signature above that. Below 200 km, this signature shifts eastward in longitude with height, and shows negligible shift above that. The meridional winds on the other hand do not show any significant longitudinal dependence at a constant local time in the thermosphere. The eastward phase tilt of zonal winds in the MLT region results in strong vertical gradients of zonal winds at certain longitudes, i.e. at around $\sim$-120$^o$E, $\sim$-40$^o$E, $\sim$90$^o$E, and $\sim$180$^o$E. These longitudes coincide with weak GW activity propagating up from the MLT region into the middle-upper thermosphere. The refraction of GWs by the background winds can induce directional dissipation (either through dynamic instability or directly via molecular viscosity). It is likely that strong vertical gradient of zonal winds due to wavenumber-4 signature dissipates upward propagating GWs by critical level filtering resulting in the similar signature in the GW activity. 

A wavenumber-4 in the thermosphere and ionosphere has been widely known to be associated with non-migrating tides. Waves that can result in wavenumber-4 component in the zonal winds include the non-migrating tides: DE3, SE2, DW5, SW6, SE2, TW7, and TE1 and the stationary planetary wave, SPW4 \cite{Forbes08, Oberheide11}. The corresponding primary GW sources include latent heat release in deep convective clouds in the tropical troposphere, net radiative heating, and wave-wave interactions at higher altitudes \cite{Hagan02, Hagan03, Oberheide07, Zhang10b}. They can propagate up into the lower thermosphere and can modulate the ionospheric E-region dynamo (100-170 km) resulting in wavenumber-4 pattern in the ionosphere at a constant local time \cite{Sagawa05, Immel06, Jin08}. Wavenumber-4 patterns have also been observed in the thermosphere at F-region heights in CHAMP observations \cite{Luhr07, Hausler09}, either due to direct penetration of waves from the troposphere \cite{Hagan09, Liu09b, Miyoshi12} or as a result of non-linear interaction between different tidal modes or between tides and planetary waves \cite{Forbes08, Hagan09, Oberheide11}. Since the model used in this study does not have a coupled ionosphere, this thermospheric signature appears independently of the corresponding signature in the ionosphere.

Amongst the above-mentioned waves, the contributions of DE3, SPW4, and SE2 to thermospheric wavenumber-4 signal are the largest \cite{Oberheide11}. Table 1 of \citeA{Oberheide11} showed that DE3, SPW4, SE2 contribute about 49$\%$, 24$\%$, 22$\%$ to wavenumber-4 signal in September at equatorial latitudes in the MLT region. Specifically, in WAM, the climatology of DE3 has been studied \cite{Akmaev08b} and shown to agree well with the SABER observations and previous observations \cite{Oberheide06, Forbes08}. Considering our model simulations and results from previous studies, it is likely that the wavenumber-4 signature in GW activity shown in Figures \ref{lt_var}, \ref{seas_var} and \ref{lon_alt} is primarily caused by the DE3 tide in WAM with significant secondary contributions from SPW4 and SE2. Further diagnoses of the tidal components in WAM GW activity will be the subject of future investigations. To the best of our knowledge, the evidence of wave-4 signature in medium-scale thermospheric GWs as shown here, is the first model result of this kind in the thermosphere.

\section{Summary and Concluding Remarks}
Wave-like perturbations have been observed in the IT region since many decades during geomagnetically quiet times \cite<e.g.,>{Evans83, Waldock86, Forbes95}, and many studies have suggested the lower atmosphere to be a significant source of these waves \cite{Waldock87, Frissell16}. Medium-scale gravity waves (GWs; of 200-600 km wavelengths) contribute significantly to the momentum budget and variability of the IT region, however their observations in the thermosphere remain far and few. In this study, we analyze the global climatology of medium-scale GWs (in terms of zonal wind variance) simulated by the high-resolution Whole Atmosphere Model (WAMT254) from the troposphere to the thermosphere for quiet geomagnetic conditions. The key findings are as follows:

\begin{itemize}

\item In the troposphere, we find that GW hotspots are confined to the well-known orographic sources (e.g., Andes, Himalayas) around the Earth. As GWs propagate up into the stratosphere, their hotspots smear out forming enhanced latitudinal band-like signatures. In the stratosphere, a larger seasonal dependence is observed, with GW activity due to both non-orographic (e.g. winter jets, fronts in the troposphere and the winter polar vortex) and orographic sources (e.g. Andes mountains, Antarctic Peninsula, Tibetan plateau). At low-mid latitudes during the warm season, effects of moist convection are observed largely during solstices. Overall, we find that the global distribution of WAM GW activity has some differences with the HIRDLS observations but captures the major global GW hotspots. 

\item In the thermosphere, the magnitudes and distribution of global GW activity in WAM is in reasonable agreement with GOCE observations at $\sim$250 km. Both show enhancements in GW activity at mid-high latitudes in the winter hemisphere during solstices. These waves are likely a combination of primary and higher-order waves originating from both orographic and non-orographic sources in the lower atmosphere. Both simulations and data also show large GW activity around the geomagnetic poles in all the seasons. Considering quiet geomagnetic conditions, this is a peculiar observation and agrees with previous observations.
 
 \item On examining the local time dependence of thermospheric GW activity, during solstices, we find a diurnal variation with minima between $\sim$15-20 LT, biased towards the summer hemisphere, and a semidiurnal variation at mid-high winter latitudes. A similar diurnal variation in GW activity was also observed by \citeA{Park23} using GRACE, GRACE-FO and CHAMP observations. Our results show that when the perturbation in only the zonal winds is considered, GWs are more likely to propagate up from the MLT into the thermosphere during the westward phase/weakly eastward phase of different tidal winds, introducing different local time variabilities in GW activity. This signifies the dominance of eastward momentum flux of GWs in the MLT, agreeing with recent results by \citeA{Liu24}. A thorough investigation into the possible mechanisms would require further targeted modeling studies but our results show migrating tides as a dominant source of local time variability in thermospheric GW activity.
 
 \item In WAM, the thermospheric GW activity exhibits a wave-4 signature when observed at a constant local time. This signature is more prominent between $\sim$6-12 LT in September, and moves eastward with local time. Our results also show that this signature closely aligns with the wave-4 signal in the zonal winds, likely indicating the impact of strong non-migrating tides in the MLT region, e.g. DE3, SE2, and SPW4. The GW activity is weaker at longitudes where the zonal winds have strong vertical gradient in the MLT due to eastward propagating wave-4 signal. This hints at critical level filtering of GWs at certain longitudes which then manifests as a wave-4 signature. Further investigation into the contribution of each non-migrating, and the validation of this longitudinal variability will be the subject of a future study. To the best of our knowledge, these are the first predictions of wave-4 signal in daytime thermospheric GW activity.

\item Analysis of medium-scale thermospheric GWs comes with inherent challenges, such as lack of global observations and need for high-resolution numerical modeling. Overall, we find a significant degree of consistency between WAMT254 and different observations and previous studies. Certain differences especially in the low latitudes of stratosphere and thermosphere arise from inherent model-data and data-data differences. For e.g., convectively forced GWs in WAM are highly sensitive to the tuning parameters of the sub-grid convective scheme. Furthermore, the numerical framework of the model, damping, and diffusion would also significantly affect the propagation and dissipation of GWs. Currently, FV3WAM with non-hydrostatic dynamical core and higher spatial resolution is being developed which will further improve the modeling of medium-scale thermospheric GWs from surface to exobase. 

\end{itemize}

\acknowledgments
This work was supported by NSF SWQU AGS 2028032. GOCE data is downloaded from \url{http://thermosphere.tudelft.nl/}. The GRACILE gravity wave data set is publicly available and can be downloaded from the PANGAEA open-access world data center at https://doi.org/10.1594/PANGAEA.879658. The WAM simulations used in this study can be downloaded from \url{https://tinyurl.com/2wjmu4vn}.

\newpage

\bibliography{papers}
\end{document}